\documentclass[twocolumn]{aastex631}
\usepackage{xspace}

\definecolor{myblue}{RGB}{0, 0, 180}

\def\SEDz*{\emph{SEDz*}}
\def\chisq{$\tilde{\chi}^2$}

\gdef\deg{$^{\circ}$}

\gdef\ltsima{$\scriptscriptstyle \; \buildrel < \over \sim \;$}
\gdef\simlt{\lower.3ex\hbox{\ltsima}}

\gdef\gtsima{$\scriptscriptstyle \; \buildrel > \over \sim \;$}
\gdef\simgt{\lower.3ex\hbox{\gtsima}}

\gdef\about{\raise.3ex\hbox{$\scriptscriptstyle \sim $}}

\def\gs{\mathrel{\raise0.35ex\hbox{$\scriptstyle >$}\kern-0.6em 
\lower0.40ex\hbox{{$\scriptstyle \sim$}}}}
\def\ls{\mathrel{\raise0.35ex\hbox{$\scriptstyle <$}\kern-0.6em 
\lower0.40ex\hbox{{$\scriptstyle \sim$}}}}

\def\Msun{\rm{\hbox{$\,$M$_{\odot}$}}}				
\def\MsunYr{\rm{\hbox{M$_{\odot}$} yr$^{-1}$}}           

\def\24m{\hbox{24\,$\micron$}$\,$}
\def\10-18{\hbox{$\times~10^{-18}$}}
\def\deg{\hbox{$^{\circ}$}}

\def\T0{{$t_0$}}

\defcitealias{Merlin2022}{Paper II}
\defcitealias{Leethochawalit2022}{Paper X}
\defcitealias{Nanayakkara2022}{Paper XVI}

\shorttitle{JWST-SEDs-Star formation histories}
\shortauthors{Dressler et al. }
\graphicspath{{./}{figures/}}

\begin{document}

\title{Early Results From  GLASS-JWST. XVII:  Building the First Galaxies -- Chapter 1. Star Formation Histories for $5<z < 7$ galaxies}

\author[0000-0002-6317-0037]{Alan Dressler}
\affiliation{The Observatories, The Carnegie Institution for Science, 813 Santa Barbara St., Pasadena, CA 91101, USA}

\correspondingauthor{Alan Dressler}
\email{dressler@carnegiescience.edu}

\author[0000-0003-0980-1499]{Benedetta Vulcani}
\affiliation{INAF Osservatorio Astronomico di Padova, vicolo dell'Osservatorio 5, 35122 Padova, Italy}

\author[0000-0002-8460-0390]{Tommaso Treu}
\affiliation{Department of Physics and Astronomy, University of California, Los Angeles, 430 Portola Plaza, Los Angeles, CA 90095, USA}

\author{Marcia Rieke}
\affiliation{Steward Observatory, University of Arizona, 933 N Cherry Ave, Tucson, AZ 85721, USA}

\author[0000-0002-6317-0037]{Chris Burns}
\affiliation{The Observatories, The Carnegie Institution for Science, 813 Santa Barbara St., Pasadena, CA 91101, USA}

\author[0000-0003-2536-1614]{Antonello Calabr\`o}
\affiliation{INAF Osservatorio Astronomico di Roma, Via Frascati 33, 00078 Monteporzio Catone, Rome, Italy}


\author{Andrea~Bonchi}
\affiliation{INAF Osservatorio Astronomico di Roma, Via Frascati 33, 00078 Monteporzio Catone, Rome, Italy}
\affiliation{ASI-Space Science Data Center,  Via del Politecnico, I-00133 Roma, Italy}

\author[0000-0001-9875-8263]{Marco~Castellano}
\affiliation{INAF Osservatorio Astronomico di Roma, Via Frascati 33, 00078 Monteporzio Catone, Rome, Italy}

\author[0000-0003-3820-2823]{Adriano Fontana}
\affiliation{INAF Osservatorio Astronomico di Roma, Via Frascati 33, 00078 Monteporzio Catone, Rome, Italy}

\author[0000-0003-4570-3159]{Nicha Leethochawalit}
\affiliation{School of Physics, University of Melbourne, Parkville 3010, VIC, Australia}
\affiliation{ARC Centre of Excellence for All Sky Astrophysics in 3 Dimensions (ASTRO 3D), Australia}

\author[0000-0002-3407-1785]{Charlotte Mason}
\affiliation{Cosmic Dawn Center (DAWN), Denmark}
\affiliation{Niels Bohr Institute, University of Copenhagen, Jagtvej 128, DK-2200 Copenhagen N, Denmark}

\author[0000-0001-6870-8900]{Emiliano~Merlin}
\affiliation{INAF Osservatorio Astronomico di Roma, Via Frascati 33, 00078 Monteporzio Catone, Rome, Italy}

\author[0000-0002-8512-1404]{Takahiro Morishita}
\affiliation{Infrared Processing and Analysis Center, Caltech, 1200 E. California Blvd., Pasadena, CA 91125, USA}

\author[0000-0002-7409-8114]{Diego~Paris}
\affiliation{INAF Osservatorio Astronomico di Roma, Via Frascati 33, 00078 Monteporzio Catone, Rome, Italy}


\author[0000-0001-5984-0395]{Marusa Bradac}
\affiliation{University of Ljubljana, Department of Mathematics and Physics, Jadranska ulica 19, SI-1000 Ljubljana, Slovenia}
\affiliation{Department of Physics and Astronomy, University of California Davis, 1 Shields Avenue, Davis, CA 95616, USA}

\author[0000-0001-9261-7849]{Amata Mercurio}
\affiliation{INAF -- Osservatorio Astronomico di Capodimonte, Via Moiariello 16, I-80131 Napoli, Italy}

\author[0000-0003-2804-0648 ]{Themiya Nanayakkara}
\affiliation{Centre for Astrophysics and Supercomputing, Swinburne University of Technology, PO Box 218, Hawthorn, VIC 3122, Australia}

\author[0000-0001-8751-8360]{Bianca M. Poggianti}
\affiliation{INAF Osservatorio Astronomico di Padova, vicolo dell'Osservatorio 5, 35122 Padova, Italy}

\author[0000-0002-9334-8705]{Paola~Santini}
\affiliation{INAF Osservatorio Astronomico di Roma, Via Frascati 33, 00078 Monteporzio Catone, Rome, Italy}

\author[0000-0002-9373-3865]{Xin Wang}
\affil{Infrared Processing and Analysis Center, Caltech, 1200 E. California Blvd., Pasadena, CA 91125, USA}

\author{Karl Misselt}
\affiliation{Steward Observatory, University of Arizona, 933 N Cherry Ave, Tucson, AZ 85721, USA}

\author{Daniel P. Stark}
\affiliation{Steward Observatory, University of Arizona, 933 N Cherry Ave, Tucson, AZ 85721, USA}

\author{Christopher Willmer}
\affiliation{Steward Observatory, University of Arizona, 933 N Cherry Ave, Tucson, AZ 85721, USA}



\begin{abstract}
JWST observations of high redshift galaxies are used to measure their star formation histories -- the buildup of stellar mass in the earliest galaxies.  Here we use a novel analysis program, \emph{SEDz*}, to compare near-IR spectral energy distributions for galaxies with redshifts $5<z<7$ to combinations of stellar population templates evolved from $z=12$. We exploit  NIRCam imaging in 7 wide bands covering $1-5\mu m$, taken in the context of the GLASS-JWST-ERS program, and 
use \emph{SEDz*} to solve for well-constrained star formation histories for 24 exemplary galaxies.  In this first look we find a variety of histories, from long, continuous star formation over $5<z<12$ to short but intense starbursts -- sometimes repeating, and, most commonly, contiguous mass buildup lasting $\sim0.5$ Myr, possibly the seeds of today's typical, $M^*$ galaxies.
\end{abstract}



\section{Introduction: Star Formation Histories: Building the First Galaxies}

The buildup of stellar mass in a galaxy is expressed as its {\it star formation history} (SFH), that is, the rate of star formation 
over an epoch of cosmic time.  Measuring the SFHs of galaxies over cosmic time has been a long-sought goal of extragalactic astronomy, extending back at least as far as the 1950s \citep{Baade1951}.  The best and most direct method, measuring the Hertzsprung-Russell diagrams of populations of stars for our own Galaxy and its nearest neighbors -- spectacularly advanced by the Hubble Space Telescope  \citep[e.g.,][]{Tosi1989, Tolstoy1996, Gallart1996,Aparicio2009, Cignoni2010} -- is unfortunately limited to a number of present-day galaxies \citep[e.g.,][]{Dalcanton2009, Dalcanton2012} in the local universe.  

To look back in time to galaxies at cosmic  distances has been critically challenging because \emph{unresolved} stellar populations cannot constrain SFHs for ages older than 2 Gyr, beyond which the dependence of an integrated spectrum on age and metallicity thwarts any comprehensive analysis \citep[e.g.,][]{Worthey1994}.  The ability to resolve SFHs over $<$2 Gyr of `lookback' owes to the rapid evolution of light-dominating A-stars over this time  period \citep{Dressler1983}, resulting in a tight correlation of SFH and spectral energy distribution (SED) that quickly vanishes for ages $\tau >$ 2 Gyr \citep[see also][]{Poggianti1999}.  The  discovery of \emph{late bloomer galaxies} -- i.e. galaxies with \emph{rising} SFRs (rather than falling as conventionally described,  \citealt{Dressler2018}) -- made use of this by observing the growth of galaxies from { $z \sim 0.75$ to $z \sim 0.45$}, a significant (2 Gyr) fraction of their lifetimes. The prevailing model for SFHs has been for homologous log-normal forms \citep[rapidly rising and slowly declining] {Gladders2013} over a Hubble time \citep{Diemer2017}, but \citet{Dressler2018} found that $\sim$25\% of Milky-Way-mass galaxies more than doubled their mass during this relatively late period in cosmic history. This study made clear that the SFHs of the first billion years of star formation, $5 < z < 20$,  would greatly benefit from the strong correlation of SED and SFH. Therefore, the potential of the James Webb Space Telescope  (JWST) to produce such data has been eagerly awaited \citep[see, e.g.,][for state of the art HST-based work]{Tacchella2022}.

In this paper we show very early results of this potential through the analysis of $5<z<7$ SEDs derived from 7-band NIRCam images obtained  as part of the GLASS-JWST Early Release Science program \citep{Treu2022a}. We start from the fluxes for $\sim$6600 galaxies extracted  by \citet[Paper II]{Merlin2022} in NIRCAM parallel observations of NIRISS spectroscopy  of the cluster Abell 2744.  We here exploit  a program written expressly for the purpose of  reconstructing the histories of galaxies at $5<z<7$ from the rich information  encoded in their SEDs, for the first time on real data. A recent paper by \citet{Whitler2022} applies similar techniques to JWST data of galaxies at $z>8$, when the universe was less than 500 Myrs old. By focusing on a sample at slightly lower redshift, we have access to redder rest-frame wavelengths and give sufficient time for older stellar populations to emerge.

These first results demonstrate that this technique will be a powerful tool in studying and understanding the buildup of stellar mass -- the  growth -- of the first galaxies in our universe. 

The paper is organized as follows: Section~\ref{sec:examples} { and  Appendix A  explain and summarize the methodology, in advance of  a forthcoming paper that will provide more detail. } Section~\ref{sec:sample} presents the observations and the selection of the sample. Section~\ref{sec:results}  shows 24 examples of the SEDs of $5 < z < 7$, 
with a focus on examples of various SFHs that we find -- mainly long and short falling SFHs, and prominent bursts.  We describe the character of these SFHs and comment on the  fidelity and robustness of the results, aided by { Appendixes B and C that also investigate the reliability of our identifications.}  Section~\ref{sec:summary} briefly contemplates the bounty expected from deeper, wider surveys, already on the JWST 
observing schedule.

A standard cosmology with $\Omega_{\rm m}=0.3$ $\Omega_{\Lambda}=0.7$ and H$_0$=70 km s$^{-1}$ Mpc$^{-1}$ is adopted. 

\section{Deriving the SFHs of the first galaxies from NIRCAM photometry}
\label{sec:examples}

The method we apply to derive SFHs follows the work of \cite{Kelson2016}, who developed a maximum-likehood   Python code to deconstruct observed SEDs from the Carnegie-Spitzer-IMACS Survey \citep[CSI,][]{Kelson2014}, in terms of best-fit stellar population templates.  The program proved particularly effective in isolating the light from younger ($<$2 Gyr) populations of A stars, confirming an early photometric study \citep{Oemler2013} that showed that 25\% of Milky-Way-mass galaxies at $z\sim0.5$ had \emph{rising} SFRs (rather than \emph{falling} at $z<1$, as conventionally described), producing at least 50\% of their stellar mass in the previous 2 Gyr -- so-called {\it late bloomers}.  

Dressler has followed this approach by writing \SEDz*, a Fortran code for analyzing stellar populations for $z>5$ galaxies, all of which have stellar ages of  $<$1 Gyr. \SEDz* was developed to answer a basic question: what do SFHs of the earliest galaxies look like? That is, its purpose is to quantify SFHs -- their forms -- for a sample of very young galaxies. Other codes measure SFHs, of course, for example, PROSPECTOR \citep{Prospector2021}, EAZY \citep{Brammer2008}, and BEAGLE \citep{Chevallard2016}.  They are essential tools in the study of \emph{galaxy evolution}: they can inform on many properties and behaviors for galaxies over a wide range of cosmic histories. \SEDz* was developed to focus on SFHs over a special epoch. It takes advantage of a unique astrophysical opportunity: the light from galaxies in the first billion years is dominated by A-stars, and A-stars are the best clock for studying young stellar populations \citep{Dressler1983, Couch1987}.  Because they evolve rapidly over a Gyr, the SFHs of A-star-dominated populations can be \emph{derived} (rather than \emph{inferred}) from SEDs,  and vice-versa. There are no complexities:  in addition to providing the proper clock, A stars are among the simplest  stars \citep{Morgan1973}; they are main-sequence stars with relatively simple atmospheres. Hydrogen absorption dominates the opacity -- not metal lines, thereby much reducing the influence of metal abundance and the environments of star formation. These are significant advantages for \SEDz* in measuring SFHs during the first billion years of cosmic history.

Operationally, the data inputs to \SEDz* are SEDs -- here, flux measurements in \emph{JWST}-NIRCam's wide bands.\footnote{F090W, F115W, F150W, F200W, F277W, F356W, \& F444W.  Fluxes were also calculated for F335M and F410M, as part of the NIRCam GTO program, but these non-independent bands were not used in the fitting here, because they are not available in GLASS-JWST.} \SEDz* uses a non-negative least squares (NNLS) engine \citep{Lawson1995}, which combines custom star-formation templates {  based on work by \cite{Robertson2010} and further developed by 
\cite{Stark2013}\footnote{ The templates are based on \citet{Bruzual2003} models, but include emission and non-stellar continuum from star forming regions in calculating the fluxes.}} (see \S3), and used by the NIRCam team to simulate observations of galaxies in the GTO deep fields \citep{Williams2018}. The custom templates made for \SEDz* cover the redshift range $3<z<12$.\footnote{No reliable templates are available at higher redshift, and the resolution in redshift is likely too fine to use \SEDz* for studying SFHs.} { For each template there is a starting redshift with fluxes proportionate to stellar mass, along with the evolution of this template, that is, the expected fluxes when observing that stellar population at subsequent epochs.}  The program divides the  $3<z<12$ epoch into integer steps, ({\it i.e.,} 12, 11, 10...3) and operates with two sets of SED templates: { one set is for 10 Myr bursts starting at epochs $5 < z < 12$, and the other is for continuous star formation  (CSF) starting at redshift as early as $z$ = 10, but not after $z$ = 4.\footnote{ Though entered as bursts, these 10 Myr episodes of star formation are indistinguishable from continuous star formation of the same mass over that epoch, as viewed from later epochs.}

The process is to build up the stellar population by combining templates, epoch by epoch. For each step NNLS works to improve the fit of combined fluxes to the \emph{observed} SED, at each step evolving the preceding star formation forward and adding the fluxes of subsequent populations, like building a wave. For example, starting with a single 10 Myr burst at $z$ = 12, \SEDz* uses NNLS to find the mass that minimizes the chi-square $\tilde{\chi}^2$ of the fit to the observed SED. Next, using the template of the $z$ = 12 burst evolved to $z$ = 11,  and adding a population starting at $z$ = 11, NNLS finds the stellar mass for each epoch (as small as zero) that further lowers the $\tilde{\chi}^2$. The process continues for bursts at 10, 9...5, and adds potential CSF populations (in units of 1.0 \MsunYr,} { over the $z$-integer-bound epoch) starting at $z$=10 and as late as $z$ = 4 -- again, as needed to improve the fit of the composite stellar populations to the observed SED. Detecting a CSF population defines the epoch of observation (OE) and ends the possibility of later bursts. To summarize, from $z$ = 12 to $z$ = 4, a sum of bursts and the possible onset of CSF improves the goodness-of-fit of model-to-observed SED, measured as reduced  $\tilde{\chi}^2$.  

\SEDz* does not prejudge the SFH.  By using a combination of individual bursts and CSF templates, \emph{SEDz*} has the flexibility to describe rising or declining, as well as bursty SFHs. Whatever the elements contributing to the best-fit to the observed SED, there is no preference or bias for any particular history. SEDz* provides error bars for each mass point in the SFH. If they are large enough, multiple histories \emph{are} possible, but this is not the case here. This means that, for a sample like this one, each SFH generated by \SEDz* is unique -- it is not meaningful to ask if another SFH would fit as well or better. With \SEDz* the minimum $\tilde{\chi}^2$ solution \emph{defines} the SFH: there are no additional indicators, or diagnostics, of whether this is the ``real" SFH, or not.\footnote{{ \SEDz* is similar to the ``non-parametric" mode of PROSPECTOR \citep{Prospector2021}, in that it does not require simplified assumptions about the functional form of the SFH that are common in other methods \citep{Chevallard2016}.}}}

In this first exploration we neglect the potential impact of dust. We will revisit the issue in future papers, when we will analyze larger samples in detail. { However, we point out that most of the mass we find in running \emph{SEDz*} on the present data is in comparatively old populations (hundreds of Myr), for which dust is not likely. In support of this, we note that galaxies \emph{in this sample} are well described by SEDs { with minimal dust, even for the youngest populations, and do not resemble the strong attenuation of ultraviolet light that is characteristic of populations with even moderate amounts of dust.  Our choice is} consistent with the results of several papers that show these initial JWST selected samples are uniformly blue \citep[Paper XVI]{Nanayakkara2022} and fairly dust free \citep[see][for a theoretical perspective]{Finlator2011,Jaacks2018,MTT2022}, except for a few rare examples \citep{Barrufet2022}.

As discussed below, the main outputs of \SEDz* are redshifts and SFHs.\footnote{SFHs derived from broad-band photometry have resolutions of tens of Myr, compared to the 10 Myr or better resolution using spectroscopic data.} To evaluate the performance in 
determining redshifts,  a simulated catalog of galaxies following the distribution of \cite{Williams2018} was prepared in the context of the JWST Data Challenge 2 (DC2) to exercise the GTO pipeline. Simulated NIRCam images of deep fields were created to test a wide range of analysis tools, with \SEDz* among them. 
Although redshift recovery is not its purpose, \SEDz* is competitive with the commonly-used photo-z codes (based on DC2 results), 
even to the extent of  identifying $z < 3$ galaxies, something \SEDz* was not designed to do and is actually outside its native ability (because stellar populations exceed an age of 1 Gyr).  

To test the performance of  \SEDz* on SFHs, a code that accepts parametric SFHs or generates stochastic SFHs was written and applied to the DC2 sample: \SEDz* was found to recover declining SFHs as well as single and multiple bursts of star formation with good fidelity, but exhibiting limited dynamic range for SFHs rising from higher redshift, as would be expected (older populations are fading while younger populations ascend). Examples of these simulations are given in Appendix~\ref{sec:app}, { along with a description of how \SEDz* is used to make these tests that provides further insight into how the program works.}

A more detailed description of the features of the code and experience in its development, along with the results of DC2 and and further SFHs tests, are planned for a forthcoming paper (Dressler et al., in prep).  \SEDz* has passed its early tests of measuring redshifts and SFHs: { the  purpose of this paper is to demonstrate its potential by applying it to some of the first real data -- from GLASS-JWST.}

\begin{figure*}
\centerline{
\includegraphics[width=7.3in, angle=0]{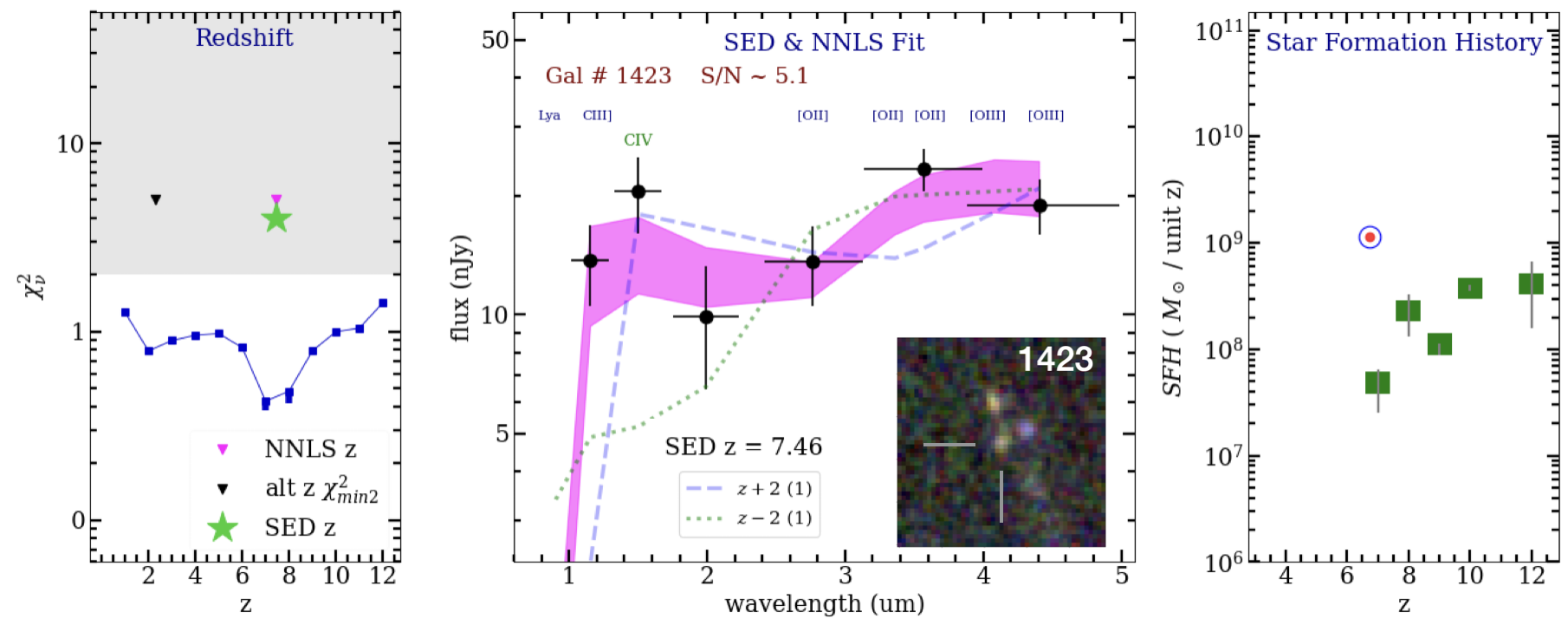}
}
\caption{
Example of main output of the \SEDz* code, run on galaxy \#1423 that has a S/N of 5.1. The three panels show the observed SED from the 7-band NIRCam imaging and NNLS fit (center), the run of $\tilde{\chi}^2$ for the fit (left), and the SFH corresponding to the best fit (right). 
{ Center:} 
The \SEDz* fit is the magenta band, showing the `quartile range' of 21 trials, with observed fluxes perturbed by 1$\sigma$ random errors.  The positions of prominent emission lines are indicated in blue labels above the SED,  with a larger green font marking a possible detection by excess flux in the band compared the best fit.  (CIV is highlighted here, though its detection is marginal.)  The dashed purple line and green dotted line show NNLS fits at a $z$ + 2 and $z$ - 2, respectively, which are the values straddling the dip in $\tilde{\chi}^2$ seen in the left panel. In the inset, the color composite image of the galaxy, based on the long wavelength camera (B=F277W, G=F356W, R=F444W), is shown. The RGB image is 3$\arcsec$ on a side.
{ Left: } The run of \chisq\ for the \SEDz* fit, with the primary minimum and a secondary minimum, if detected, marked by magenta and black triangles, respectively.  The preferred redshift, marked with a green star, corresponds to the prominent drop in \chisq\ at $z\sim$ 7. 
{ Right:} SFH for this NNLS solution. Green boxes mark the mass contributed at each epoch. The bars are quartile errors in mass that arise from the random flux errors of the 21 models.  Systematic errors in general should be comparable in magnitude to random errors and unlikely to significantly change SED shape and, with it, the SFH (see text). The red dot in the blue circle marks the mass accumulated through the epoch of observation. 
\label{fig:1423ED}}
\end{figure*}

\section{Observations and data sample}\label{sec:sample}

We use the NIRCam data obtained on June 28-29 2022 in the context of the GLASS-JWST-ERS survey \citep{Treu2022a}. They consist of images centered at RA$=3.5017025$ deg and Dec$=-30.3375436$ deg and taken in 
 seven wide filters: F090W, F115W,
F150W, F200W, F277W, F356W, and F444W.  The
5$\sigma$ depths in a $0\farcs30$-diameter aperture are in the range
29-29.5 AB mag.  The final images
are PSF-matched to the F444W band. 

Details of the NIRCam data quality, reduction, and photometric catalog creation can be found in \citetalias[]{Merlin2022}. 
As in other papers of this focus issue, we neglect the potential impact of lensing magnification, which is expected to be modest at the location of the NIRCam parallel field. We will revisit lensing magnification in future work. However, we note that the effect is small, and that lensing cannot affect SFH shapes, only their magnitude. 

Out of the  6590 detections, we select sources with $S/N > 5$. { This S/N is obtained by averaging the S/N of three bands: F200W, F277W, F356W. }
We run \SEDz* on this sample { and obtain a redshift for all the sources. From now on} we consider only  galaxies with  $5<z<7$, for a total of 123 objects.\footnote{Galaxies at $z>7$ are discussed in the companion papers by \citet{Leethochawalit2022}, \citet{Castellano2022}, \citet{Yang2022}, \citet{Santini2022} and \citet{Treu2022b}.} We visually inspected the photometry of these galaxies, removing those with artifacts or chip boundaries affecting them in at least one band. We also selected only galaxies with an \SEDz* fit ($\tilde{\chi}^2<2$), for a total of 43 galaxies. { We further reject 10 galaxies with clear (single-band) photometry errors, poor fits of the SED, or ambiguous \chisq\ (redshift) measurements.  From the remaining 33 we select 24 that best illustrate the distribution and diversity of SFHs we find.\footnote{{ Of the 9 galaxies not shown, 7 were equivalent to the burst examples of Fig. 4, and one each resembled the samples of Fig. 3 and 5. A statistical assessment of the frequency of each SFH shape requires a larger sample and is left for future work. Those galaxies are listed in the Table~2.} These are used in the following section, to demonstrate the strength of the SFH reconstruction and the variety we find.}}

\begin{table*}
\caption{Summary of the main galaxy properties of the sample discussed in the paper.}
\centering 
\begin{tabular}{lcccccccl}
\hline
\hline
  \multicolumn{1}{c}{ID} &
  \multicolumn{1}{c}{RA} &
  \multicolumn{1}{c}{DEC} &
  \multicolumn{1}{c}{z} &
  \multicolumn{1}{c}{F444w} &
  \multicolumn{1}{c}{$\log(M_\ast)$}  &
    \multicolumn{1}{c}{Notes} \\
  \multicolumn{1}{c}{} &
  \multicolumn{1}{c}{(J2000)} &
  \multicolumn{1}{c}{(J2000)} &
  \multicolumn{1}{c}{} &
  \multicolumn{1}{c}{($\mu$Jy)} &
  \multicolumn{1}{c}{($M_\odot$)} &
    \multicolumn{1}{c}{} \\
  1423 & 00:14:03.29 & -30:21:46.7 &  7.5  & 0.019$\pm$0.003 & 9.1\\
\hline
  733 & 00:14:01.51 & -30:22:12.6 & 6.5 & 0.031$\pm$0.004 & 9.5 & L\\
  2199 & 00:14:04.77 & -30:21:14.8 & 6.9 & 0.046$\pm$0.003 & 9.5 & L\\
  3459 & 00:14:01.05 & -30:17:41.7 & 5.7 & 0.032$\pm$0.004 & 9.4 & L\\
  3665 & 00:14:00.96 & -30:18:00.8 & 5.2 & 0.019$\pm$0.003 & 9.0 & L\\
  4761 & 00:13:58.16 & -30:19:23.5 & 6.4 & 0.094$\pm$0.005 & 9.7 & L, (2) \\
  5429 & 00:13:58.11 & -30:19:03.9 & 7.0 & 0.023$\pm$0.004 & 9.2 & L\\
  5879 & 00:13:54.26 & -30:18:51.8 & 5.1 & 0.018$\pm$0.003 & 8.9 & L\\
  6155 & 00:14:02.99 & -30:18:47.8 & 6.8 & 0.012$\pm$0.003 & 9.2 & L\\
  4144 & 00:14:02.29 & -30:18:31.9 &5.9& 0.024 $\pm$0.004 &9.2& L, * \\
\hline
  483 & 00:14:05.72 & -30:22:22.5 &6.5& 0.036 $\pm$0.004 &9.2& B, *\\
  1663 & 00:14:04.39 & -30:21:38.5 & 5.6& 0.007$\pm$0.002 &8.3& B, * \\
  2104 & 00:14:02.85 & -30:21:19.3 &5.6& 0.037$\pm$0.003 &9.1& B, *\\
  2321  & 00:14:06.58 & -30:21:18.2 &5.8& 0.014$\pm$0.003 &9.0& B, * \\
  2626 & 00:14:04.50 & -30:20:58.6 &6.2& 0.056 $\pm$ 0.004 &9.1& B, *\\
  2827 & 00:14:03.49 & -30:20:51.0 & 5.6 & 0.013 $\pm$ 0.003& 8.7 & B, *\\
  2850 & 00:14:03.02 & -30:20:49.0 & 5.9 & 0.026$\pm$0.004 &9.3 & B\\
  2993 & 00:14:00.58 & -30:20:40.8 & 6.7 &  0.021$\pm$0.003& 9.1 & B\\
  3824 & 00:13:53.63 & -30:18:09.7 & 6.6 & 0.019$\pm$0.003 & 9.0 & B\\
  3918 & 00:14:02.51 & -30:18:15.2& 5.5 & 0.016$\pm$0.003 & 8.8 & B \\
  4026 & 00:13:52.80 & -30:18:21.8 & 5.7 & 0.198$\pm$0.008 & 9.8 & B, (2)\\
  4587 & 00:14:02.65 & -30:19:16.2 & 5.7 & 0.028$\pm$0.004 & 9.1 & B\\
  4736 & 00:14:02.01 & -30:19:22.6& 6.3 & 0.012$\pm$0.002 & 8.6 & B \\
  5394 & 00:14:00.89 & -30:19:04.7 &6.3& 0.076$\pm$0.005 &9.3& B, *\\
  6324 & 00:14:01.78 & -30:18:13.3 & 6.6 & 0.054$\pm$0.004 & 9.4 & B\\
\hline
  680 & 00:14:05.63 & -30:22:14.9 & 5.0 & 0.02$\pm$0.003 & 8.9 & S\\
1105 & 00:14:01.61 & -30:21:58.6 & 5.4  & 0.077$\pm$0.002 & 8.6 & S\\
  1127 & 00:14:02.43 & -30:21:57.6 & 5.6 & 0.009$\pm$0.002 & 8.8 & S\\
  1356 & 00:14:04.14 & -30:21:48.7 & 5.0 & 0.017$\pm$0.003 & 8.6 & S\\
  2006 & 00:13:59.17 & -30:21:23.8 & 5.7 & 0.013$\pm$0.003 & 8.8 & S\\
  2838 & 00:14:02.83 & -30:20:48.6 & 6.7 & 0.13$\pm$0.005 & 9.6 & S,  (1)\\
  2864 & 00:14:05.89 & -30:20:47.9 & 6.4 & 0.062$\pm$0.005 & 9.4 & S, (2)\\
 2872 & 00:14:03.49 & -30:20:51.0 &5.6& 0.013$\pm$0.002 & 8.7 & S\\
 4455 & 00:14:00.00 & -30:19:10.3 & 5.1 & 0.036$\pm$0.006 & 9.2 & S, *\\
\hline
\end{tabular}
\tablecomments{L: Galaxies shown in Fig. \ref{fig:ExtendedSFHs}. B: Galaxies shown in Fig. \ref{fig:BurstySFHs}. S: Galaxies shown in Fig.~\ref{fig:short_SFHs}. (1) Galaxy also in the sample selected in \citetalias{Leethochawalit2022}; estimated redshifts are consistent within the uncertainties.  (2) Galaxies also in the sample of \citetalias{Nanayakkara2022}; estimated redshifts are consistent within the uncertainties. *: Galaxies not shown in the Figures.} 
\label{tab:gals}
\end{table*}

\section{First results: Sample SFHs of the first galaxies.}\label{sec:results}

In this section we investigate a sample of SEDs from the JWST-GLASS NIRCam observations. 
By way of example, Figure~\ref{fig:1423ED} shows the 
SED of \#1423 from the GLASS catalog of \citetalias{Merlin2022}.
This galaxy has a redshift of z = 7.5, so it does not enter the sample, but it is { otherwise indistinguishable from the 33 galaxies we have chosen, and a good example to show} the potential of \SEDz* on a wide range of observations. 

The data for \#1423 are shown in 3 panels, the format for 
\SEDz* results. The central panel shows the SED, crosses marking the fluxes 
with 7 NIRCam `wide' filters in nJy, with errors.  \SEDz* has been run as described in the previous 
section on this SED, yielding the reduced  $\tilde{\chi}^2$ distribution shown in  the left panel, with a well defined minimum at z $\approx$ 7.5, between the two lowest points. Although \SEDz* calculates the SED fit in integer redshifts, there is sufficient precision in the fits to interpolate between adjacent values; { the first decimal in the redshifts is significant.} Returning to the middle panel, the magenta band that `connects' the points is generated by randomly perturbing the input SED { with 1$\sigma$ flux errors (typically a few nJy)} and repeating the NNLS fitting 21 times.  The { width of the band} represents the quartile range of NNLS fits with this relatively low-S/N example.  The run of \chisq\ presented in the left panel shows a significant drop between $z=6$ and $z=7$, interpolated by the program to derive the $z = 7.5$ value recorded in the central panel, in reasonable agreement with the estimate of the redshift as $6.9$ obtained with the code \texttt{EAzY} \citep{Brammer2008} with the default V1.3 spectral template and a flat prior in \citetalias{Leethochawalit2022}.  

The real payoff for these data, and our methodology, is the SFH shown in the right panel: the green boxes record the stellar mass added to the galaxy at succeeding epochs.  Error bars for these boxes do not include probable systematic errors (for example, the errors associated with photometry at these faint levels), but the relative uniformity of this clearly declining rate of star formation suggests they are not severely underestimated. As we will see, this slowly falling SFH, starting at z=12 (and likely earlier), shows a combination of features we find in this early study, both parametric and stochastic.  The total mass of $10^9$ \Msun\ is shown at $z = 7.5$, the epoch of observation (OE) marked by the red dot/blue circle, while the squares showing the mass buildup are at their integer redshifts, by necessity. The galaxy  shows a factor of $\sim3$ buildup in stellar mass added after its formation at $z\sim12$.
 This relatively low mass is consistent with the 10-20 nJy fluxes measured for standard mass-to-light ratios \citep[Paper XI]{Santini2022}.


\begin{figure*}
\center
\includegraphics[scale = 0.48]{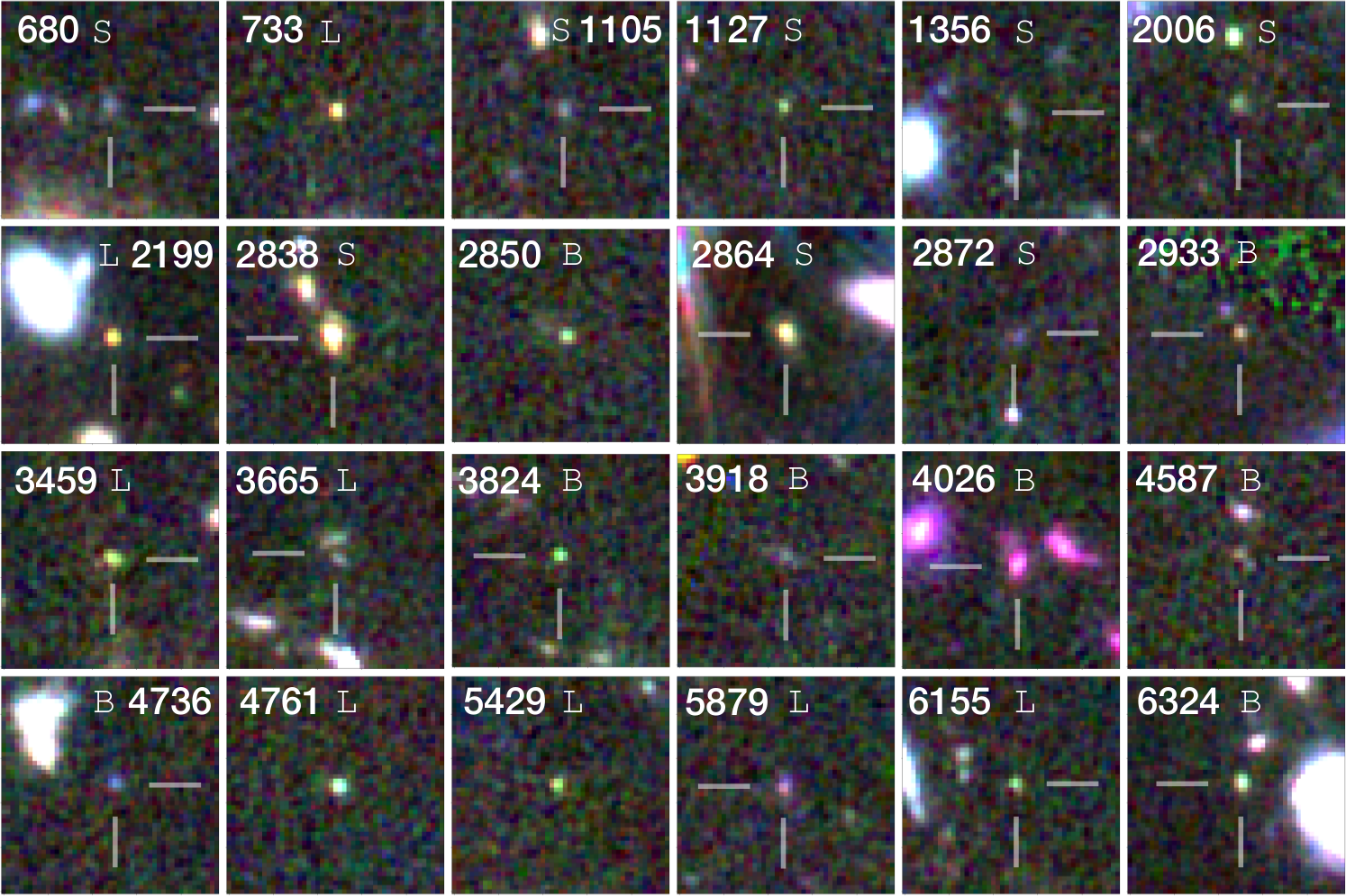}
\caption{Color composite image based on the long wavelength camera (B=F277W, G=F356W, R=F444W) for the { 33 galaxies discussed in detail in}  this paper. Individual images are degraded to the lower resolution of F444W. Postage stamps are $3\arcsec$ on a side. Pixels are 63 mas. The letter next to ID identifies the main characteristic of the SFH: S= short, L= long, B= bursty.}
\label{fig:cutouts}
\end{figure*}

We continue by showing the SEDs of 24  galaxies chosen to illustrate  the full sample.  
Details of such galaxies are given in Tab.\ref{tab:gals}, while color composite cutouts are presented in Fig.\ref{fig:cutouts}.

Figure~\ref{fig:ExtendedSFHs} shows a selection of 8 galaxies at $5<z<7$ that display long, often continuous star formation back to $z\sim11-12$ -- early times in galaxy building. Unsurprisingly, then, these SFHs are either level or declining, remarkably steady for such an early birth and growth. { Their total mass is typically a few billion} \Msun, { with only one (\#4761) approaching $10^{10}$ \Msun.} { Though they resemble the smooth, settled SFHs of much older galaxies, these SFHs are not the most common in our sample accounting for only 27\% of our sample.}

These galaxies with extended SFHs straddle the end of reionization \citep{Adam2016, Fan2006}, a full Gyr since the probable beginning of galaxy formation and growth, and appear to have been forming stars continuously from $z\sim12$.  There are indications of abrupt, order-of-magnitude, changes in stellar mass added (more and less), but mostly they are smooth SFHs.  

In the  sample of { 33 galaxies, there are no \emph{extended},  \emph{rising} SFHs.  However, this} is consistent with the results of our simulations of stochastic SFHs (Appendix A), which shows the difficulty of detecting an old fading population behind the light of an ascending, younger population \citep[e.g.,][]{Iyer2020, Whitler2022}. { On the other hand, many of the falling SFHs of histories with shorter runs of contiguous star formation (see Fig. 5) could be masking rapidly rising star formation prior to their detection.}  Larger and deeper samples will be needed to characterize such behavior and, in general, the \emph{distribution} of SFHs types, through what will essentially be continuity constraints. 

A more thorough inspection of the SEDs themselves in Figure~\ref{fig:ExtendedSFHs} reveals that all have 
a substantial older stellar population, in agreement with the derived SFHs.   Also consistent is the absence of strong emission lines (raising the flux in a wide band) that has been seen in some of the stochastic or burst-dominated cases we look at next.  


\begin{figure*}
\centering
\includegraphics[scale = 0.7]{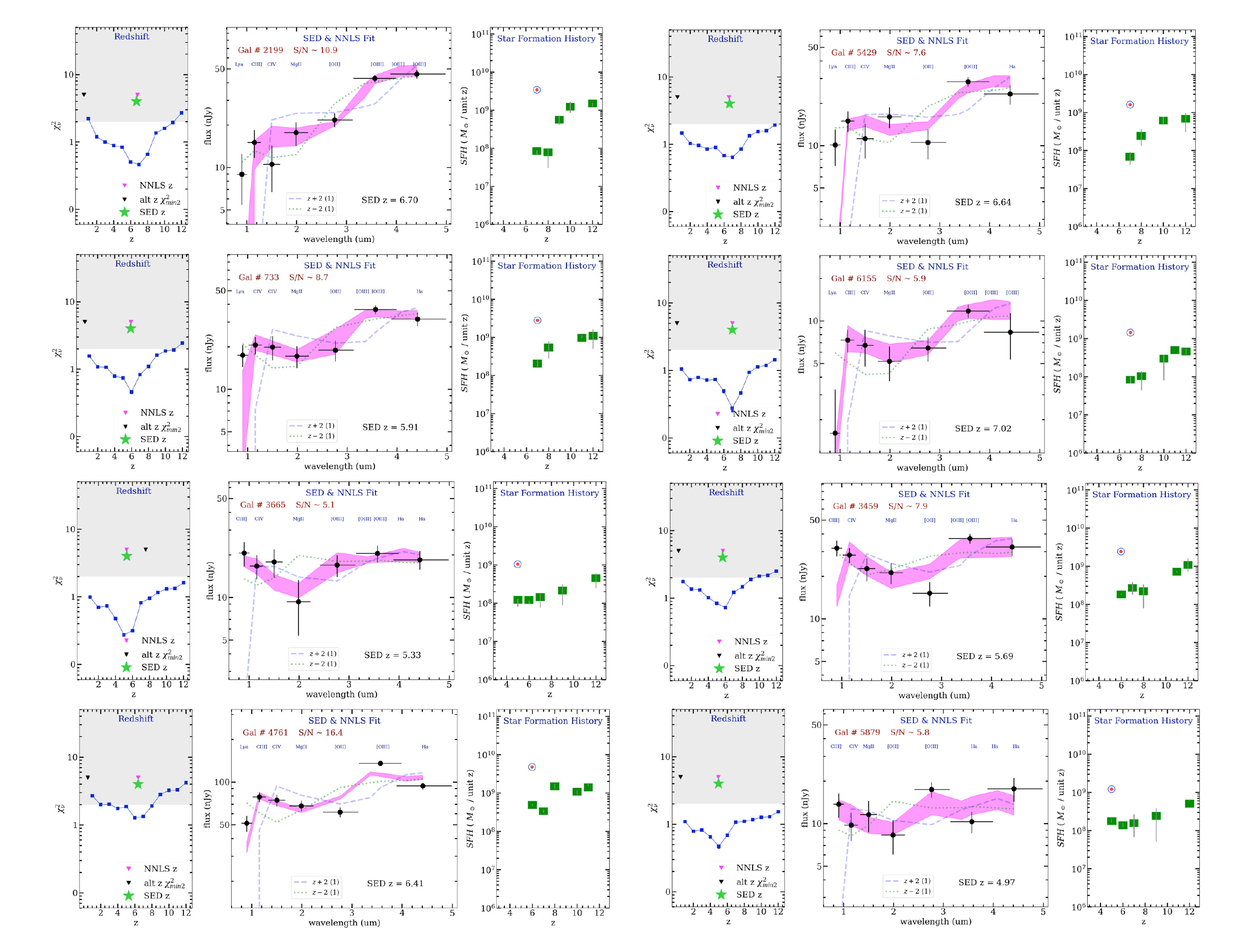}
\caption{Examples of galaxies 
dominated by long, often continuous SFHs, extending back to $z=11-12$, either level or declining. There are several cases of order-of-magnitude changes in stellar mass added (more or less), but most of the data are consistent smooth SFHs. These are not the most common SFHs in our sample: \#4144, not shown, is only addition example. Panels, lines and symbols are as in Fig.\ref{fig:1423ED}.}
\label{fig:ExtendedSFHs}
\end{figure*}

Figure~\ref{fig:BurstySFHs} shows examples of galaxies with what look like more stochastic SFHs -- bursty, repeating, and non-continuous episodes of $<200$ Myr. The top { two cases are a single, and perhaps an extended single-burst.  
The third and fifth examples are galaxies with widely separated bursts of star formation. Each} formed roughly half of their stellar mass at $z$=11-12, then experienced an comparatively { quiescent period a comparable burst of star formation was added near OE -- possibly these are interrupted versions of the long SFHs of Fig. 3.  The blue upturn of each SED relative to longer wavelengths is representative of the a younger, ongoing star formation added to a much older population, with little in-between.  

The fourth and sixth examples are two nearly identical triple-bursts (with a mass difference of a factor of 3) that look like (at least for the two pairs of early bursts) independent events, possibly followed by decaying star formation for each at the following (adjacent) epoch. The strong resemblance of the SEDs and SFHs for these two cases demonstrates the strong SED $<->$ SFH relation for $5<z<12$ galaxies that \SEDz* exploits. The last two panels show two SFHs that suggest extended star formation episodes -- 3 and 4  starbursts spread over a few hundred Myr, continuous or episodic.  These might be cases of extended-burst histories, but, according to the tests done for Figs. 8 and 9 of Appendix B, these could possibly be \emph{single} early single bursts, $z=9$ and $z=10$ that are badly resolved by the \SEDz* fitting of the SED.  Arguing against this interpretation is star formation at OE in both cases, unlike the typical simulated cases in Fig. 8.}


\begin{figure*}
\centering
\includegraphics[scale = 0.7, trim = 2.5cm 0 0 0]{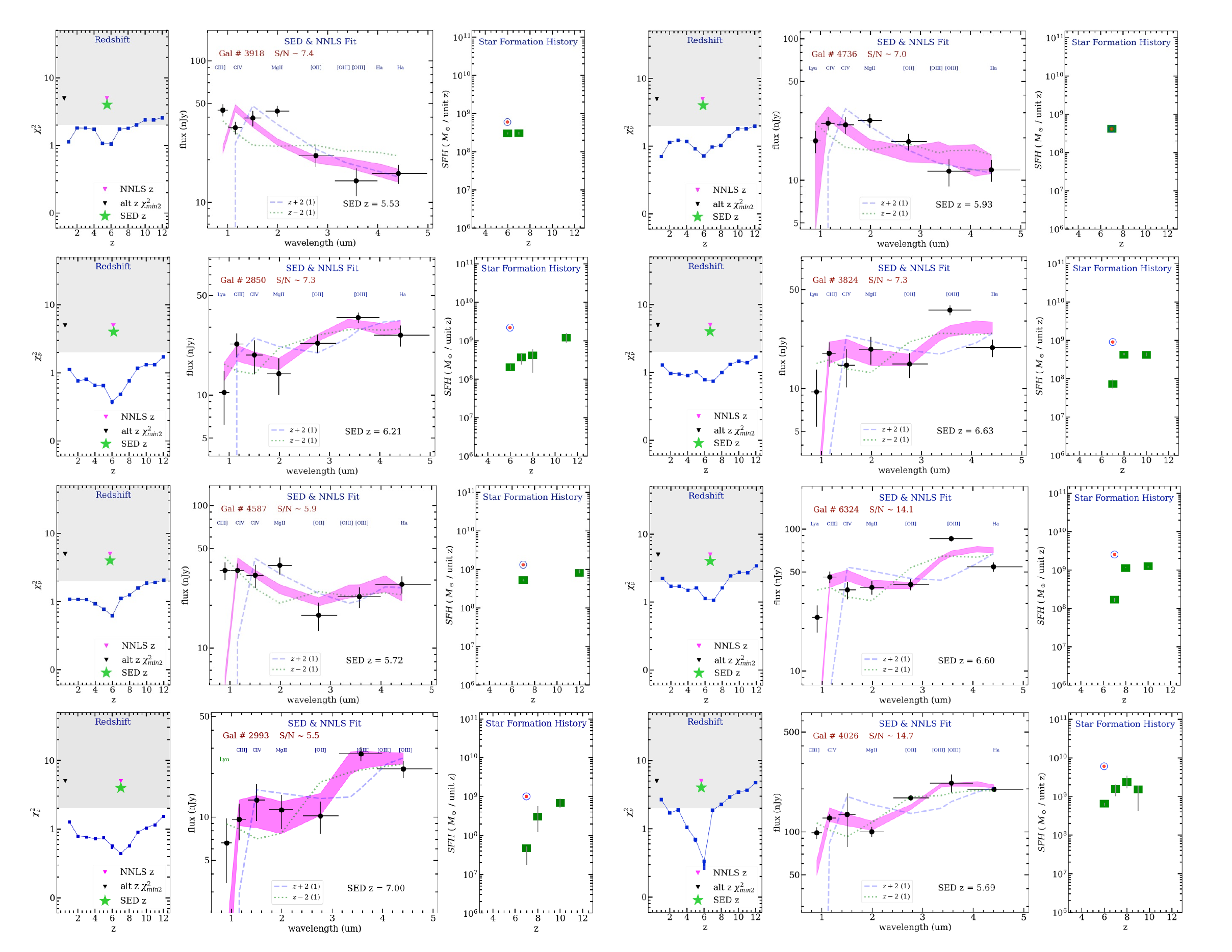}
\caption{Examples of galaxies characterized by bursty or widely-separated episodes of star formation.  The star formation episodes of the top two appear to be single bursts of $\sim5\times10^9$\Msun: 5 others of the 33-galaxy sample resemble these. Further examples appear to be multiple bursts that could be associated (see text). Panels, lines and symbols are as in Fig.\ref{fig:1423ED}.
\label{fig:BurstySFHs}}
\end{figure*}

Lastly, Figure~\ref{fig:short_SFHs} shows further diversity in high-redshift SFHs. All 8 are examples of a `run of star formation' over three adjacent epochs, typically at $z=8-7-6$. { 
As just discussed, a detection by} \SEDz* of adjacent episodes of star formation at higher redshift, observed in a galaxy at much lower redshift (e.g., {z=5 or 6}) may only be a failure to \emph{resolve} a single, larger burst. { Unlike those, however, for 6 of the 8 cases we show in Fig. 5, star formation is spread over \emph{lower} redshifts.  The epochs of 8, 7, and 6 have durations} of 125, 171, and 244 Myr over which time the SED changes markedly, { which means that their SFHs are not easily confused with bursts, for example. Rather,} these half-Gyr runs of star formation producing between $5\times10^8$ and $5\times10^9$\Msun, by redshifts 5 to 6 could be the beginnings of typical $M^*$ galaxies. The lack of rising star formation before these usually declining, short SFHs is interesting, since at these epochs lower rates at $z\sim9$, for example, should be detectable.  A possibility is that the first episode of these short SFHs is a burst lasting only tens of Myr, unresolved in the $\sim$100 Myr of the first epoch.

The presence of both short and long continuous runs of star formation, along with a diversity of starbursts -- also occurring over short or long timescales, suggests that the starburst examples are undergoing significant merger events, while the continuous SFHs, whether short or long, suggests galaxies left largely undisturbed through their early growth. 

{ We note that given the heterogeneous nature and varying quality of this early sample, we have not tried to estimate the frequencies of the different types of SFHs have identified. However, for what it's worth, the three types of SFHs in Figs. 3, 4, and 5, including the cases in the 33-galaxy sample that are not shown, amount to 27\%, 45\%, and 27\%, a roughly uniform distribution, with bursty behavior leading the way.}


\begin{figure*}[h!]
\centering
\includegraphics[scale = 0.7, trim = 3cm 0 0 0 ]{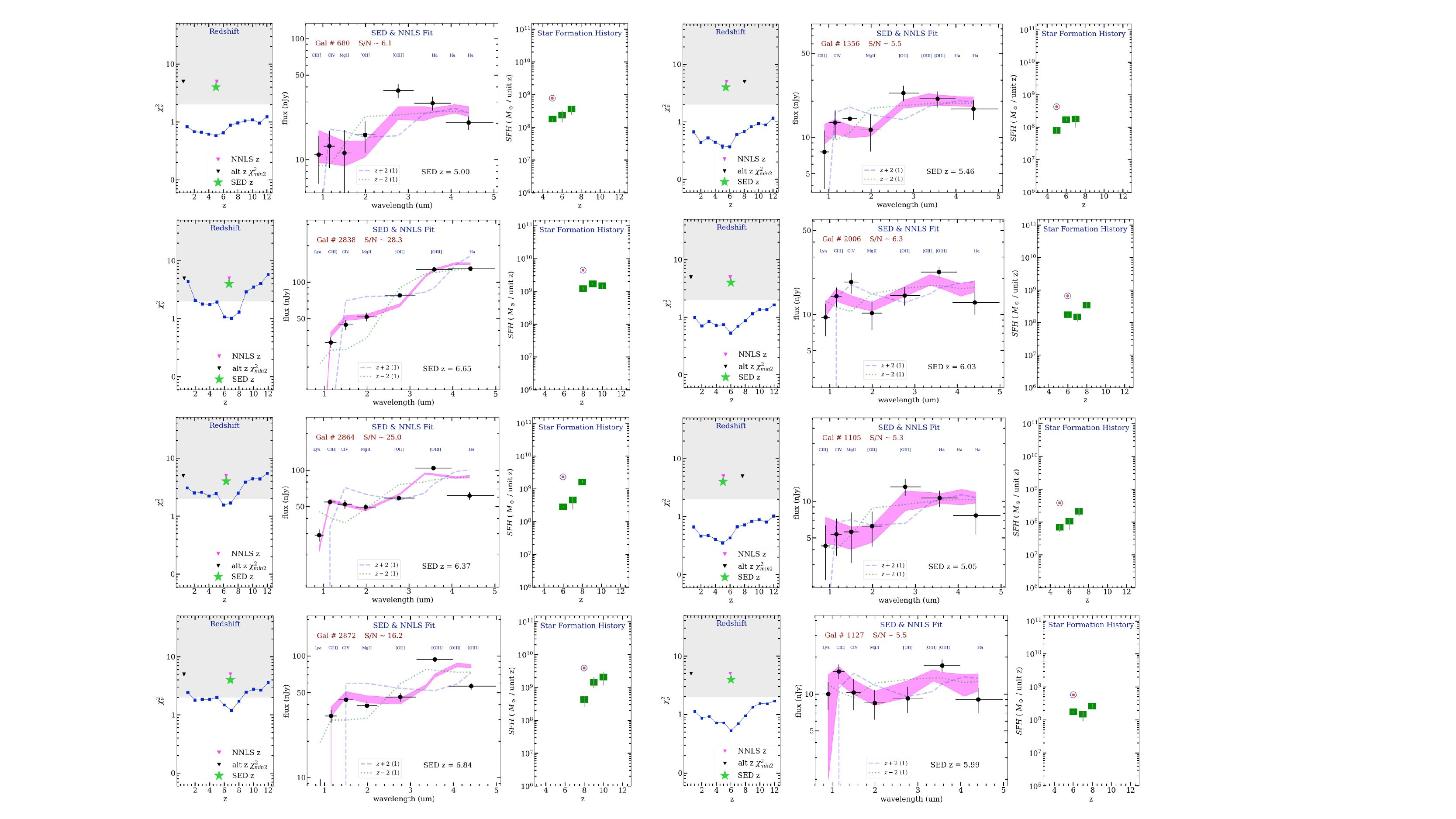}
\caption{ \label{fig:short_SFHs}
Examples of galaxies characterized by the most common type of SFHs found in this early study -- runs of star formation over 3 contiguous epochs: this might be how the most common $M^*$ galaxies are born and grow. Panels, lines and symbols are as in Fig.\ref{fig:1423ED}.}
\end{figure*}

\section{Summary and Future Work}
\label{sec:summary}

The principal goal of this paper has been to show how broad band imaging with NIRCam on JWST can provide an excellent measurement of both redshift and star formation history -- the latter is critical to understanding the assembly of the
earliest galaxies. High-redshift SFHs will also provide an additional probe of galaxy populations at $z=11-12$ and possibly beyond: the SFH of each lower-redshift galaxy probes earlier times, adding to what we will learn from actually \emph{observing} galaxies at that time.

{ Future data will include deeper NIRCam imaging with more multiple, highly-dithered images that will minimize cosmetic defects and improve photometric measurements: uniformity in all bands is critically important for SED analysis.  Improvements in signal-to-noise and defect-free imaging will be crucial for optimizing the ability of \SEDz* to delineate and distinguish SFHs for the first galaxies.}

This initial, small sample appears to exhibit the types of SFHs that astronomers would expect, including falling SFHs (``tau models") leading to the near-constant star formation rate that dominates the epoch of greatest growth, z $\sim3$ $\rightarrow 1$. But even this small sample suggests that starbursts (stochastic histories) and as-yet undetected
\emph{rising} SFHs -- expected to  dominate as we look further back to  ``the beginning'' -- are showing up.  If the results of this study are indicative of what is to come, astronomers can look forward to the James Webb Space Telescope fulfilling its ``prime'' mission better and sooner than we could have hoped.

\begin{acknowledgments}

AD gratefully acknowledges the support of the NIRCam team for the opportunity to contribute to the NIRCam science program on the earliest galaxies.  The DC2 ``deep field" simulation was a Herculean task
that provided the essential guidance for the program described here, and for the NIRCam GTO program to come.   AD also thanks the GLASS team 
for allowing him in, to lift the curtain on the growth of the first galaxies -- what the ``HST \& Beyond" Committee \citep{Dressler1986} dared to dream.

This work is based on observations made with the NASA/ESA/CSA James Webb Space Telescope. The data were obtained from the Mikulski Archive for Space Telescopes at the Space Telescope Science Institute, which is operated by the Association of Universities for Research in Astronomy, Inc., under NASA contract NAS 5-03127 for JWST. These observations are associated with program JWST-ERS-1324. { The JWST data used in this paper can be found on MAST: http://dx.doi.org/10.17909/fqaq-p393. } We acknowledge financial support from NASA through grant JWST-ERS-1324. 
D.S., K.M., M.R. and A.D. are supported by JWST/NIRCam contract to the University of Arizona, NAS5-02015.
A.M. acknowledges financial support through grants PRIN-MIUR 2017WSCC32.

\end{acknowledgments}


\appendix

\section{Testing \emph{SEDz*} with Simulated SFHs}
\label{sec:app}

{ In this Appendix we show examples of simulations of SFHs that have been used to test the robustness of the resulting SFH.   We first show how the basic types of SFHs we find can be validated by the simulation process, by examining six ``generated" examples. In Appendix B we address the uniqueness of the \SEDz* solutions, 
specifically, the robustness of the long SFHs of Fig. 3 and the possibility that some or all of these could actually be bursty SFHs misidentified by \SEDz*. 

\SEDz* solves for the buildup in stellar mass per epoch by finding a maximum-likelihood solution of the observed SED from combinations of stellar population templates. The process is \emph{reversible} because of the strong covariance of SFH and SED over the first Gyr of cosmic history. Our tests of \SEDz* consists of `composing' SFHs, like the green squares in Figs. \ref{fig:ExtendedSFHs}, \ref{fig:BurstySFHs},  \ref{fig:short_SFHs} over $5<z<12$ and then using \SEDz* ``backwards", to construct the SED that generates the invented mass profile.  This has been done for both parametric SFHs, such as smooth rising or falling SFHs over many epochs, shorter connected epochs of star formation, or bursts. The three steps are: (1) we invent SFHs like those observed in the figures; (2) use the  \SEDz* code in reverse to produce the SED that would generate the simulated SFH; and (3) use \SEDz* to generate 21 SFHs by perturbing the SED by expected random errors, including error bars on the green boxes to represent the quartile ranges of those 21 solutions. Comparing the simulated SFH and the recovered SFH tests the fidelity of the code.  }

\begin{figure*}
\includegraphics[scale = 0.9, trim = 10cm 0 0 0cm]{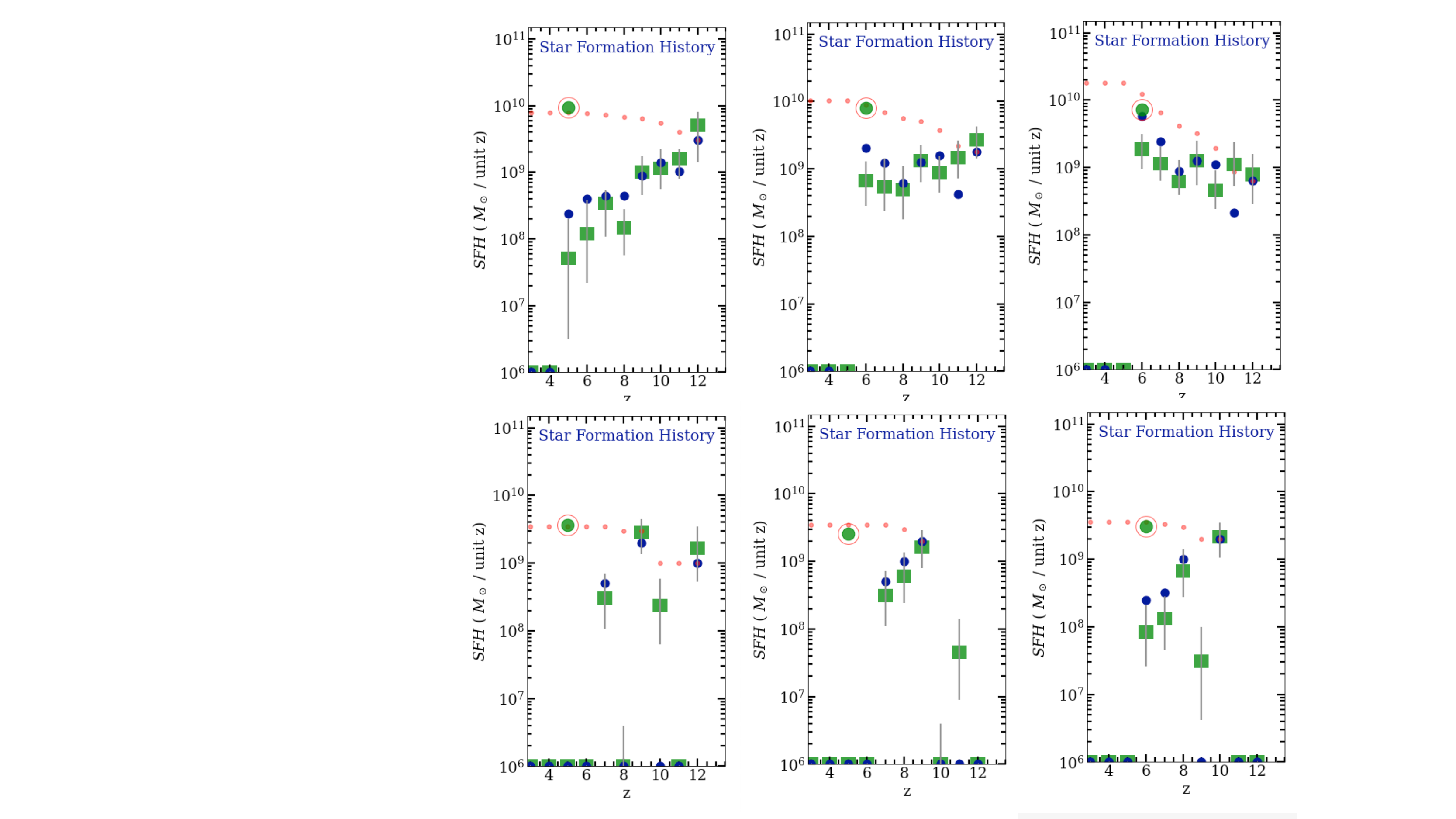}
\caption{Validation of the \emph{SEDz*} code through simulations. Each panel shows a different input SFH (dark blue dots) and how the code retrieves it (green squares) through SED fitting. Red dots show the integrated mass at each step with a green-dot/red-ring marking the total mass.  Details on the characteristics of the different input SFHs are given in Appendix~\ref{sec:app}.
\label{fig:SFH_sims}
}
\end{figure*}

Figure \ref{fig:SFH_sims} shows { six examples: the dark blue dots are the `invented' run of stellar mass and the small red dots are the integrated mass, and a green-dot/red-ring for total accumulated mass.} In all cases the simulated mass buildup is well traced by the recovered mass growth, even as the epoch-to-epoch agreement bounces around, as would be seen in the evolution of this diagram stepped through earlier epochs than those shown in this Appendix and Figure~\ref{fig:SFH_sims} { (We show examples of this epoch-by-epoch buildup in Fig. 7).  }

The top three histories were constructed to resemble those of Figure 3: long, falling,  mostly continuous runs of star formation since $z=11$ or $z=12$.  We show, top left to right, continuous star formation until $z=6$ or $z=5$. In these cases the star formation is added in 10 Myr bursts only -- there is no constant star formation at OE.\footnote{In the normal running of \emph{SEDz*} bursts serve as the normal way of adding mass to the galaxy: only at OE is the burst resolvable into a component of relatively young stars, but by the next epoch the burst cannot be distinguished from constant star formation over the previous epoch, and is suitable for building up a old stellar population.}  The example on the right was designed to be smooth and declining, with a stochastic `noise' of a factor of $\sim2$ -- similar to the steeper examples in Figure 3.  The central panel of Figure 3 shows a slightly rising SFH which is recovered as slightly falling, and for the right example, the modestly rising SFH that is simulated is recovered as flat. In summary, the basic character of a long stretch of star formation is recovered, but  slopes are not accurate. This is unimportant for our purposes here, but it suggests that it might be necessary to see what systematic error produces the effect, and possibly to adjust for it.

{ The qualitative agreement is good for these top-row examples -- the smoothness and falling character is recovered, although in each case the SFH is higher earlier, and lower later, than what was simulated. The typical scatter of a factor-of-two comparing model (purple points) and recovered masses (green square) are remarkably good, and certainly sufficient for a basic description of the SFH -- the goal of this program.

The bottom panels are simulations of different degrees of bursty behavior. The bottom left case simulates 3 }independent bursts, $10^9$ \Msun\ at $z=12$, $3\times10^9$ \Msun\ at $z=9$, and $6\times10^8$ \Msun\ at $z=7$. The agreement is good, except for the alias mass added at $z=10$.  These simulations suggest that SFHs dominated by bursts, like those in Figure 4, are well recovered by \emph{SEDz*}. { Again, the displacement of the points from the model by factors-of-two are insignificant our purpose.}

The  example in the central panel is for the kinds of shorter but continuous runs of star formation shown in Figure 5.  Again, the agreement is qualitatively very good, again with a slight systematic error in slope.  Another alias is seen $z=11$, well before the simulated SFH.

Finally, the right bottom example was for a simulation of four episodes of star formation that are disjoint, but likely associated: $2\times10^9$ \Msun\ at $z=10$, $8\times10^8$ \Msun\ at $z=8$, and two at $\sim10^8$ \Msun\ at $z=7$ and $z=6$.  Again, there is a slope error, and scatter, and an alias at $z=9$, but the character of the SFH is recovered successfully.

Our more extensive simulations, like those in Fig. 6, show that \SEDz* are successful at recovering the basic form of a wide variety of SFHs.  With more experience and understanding of how to optimize this kind of testing,  and a richer, deeper sample than is now available,  \emph{SEDz*} should be able to characterize the \emph{distribution} of SFHs and perhaps even how the distribution itself evolves over the epoch $5<z<12$.

\section{How \emph{SEDz*} builds the 'best' SFH}
\label{sec:appb}

The purpose of this section is to use simulations to further illustrate how \SEDz* computes a SFH by combining fluxes from stellar population templates to find the best fit to an observed SED.  Specifically, we show that \SEDz* finds a \emph{unique} solution that is the best measurement of the SFH, and related uncertainties, free of preconceptions or bias of what that SFH might be.

We begin this section by exploring SFHs, epoch-by-epoch, of galaxies shown in Fig. 3 that we described as having long and more-or-less continuous star formation for the $\sim1$~Gyr $z=12$ to OE ($5<z<7$).  In Figure 7 we show in 3 columns 3 examples from Fig. 3 -- GLASS-5429, GLASS-3665, and GLASS-3459. For each star formation begins at or before $z=12$ and accumulates by OE to $\sim1-2\times10^9$\Msun.~ As we described in \S2, \SEDz* uses two sets of stellar population templates, for 10 Myr bursts and CSF, with fluxes recorded at integer redshifts $z=12$ to $z=3$. For each epoch, these ``template tables" 
give the fluxes expected from 1 \MsunYr\ at that redshift, and the subsequent fluxes for that evolving population. 

Each of the three tests begins with the bottom panel, with OE at $z=11$, and works up to an OE of $z=6$ (left case) or $z=5$ (center and right cases). Using GLASS-5429 (left) as an example, and starting at the bottom, \SEDz* has attempted to fit the observed SED (OE, $z=6$) with only stars born at $z=12$ and $z=11$.  The best fit is for a mass of $9\times10^8$ (right plot) at $z=12$, and no $z=11$ contribution, producing an SED (magenta line -- center panel) that is a very poor fit to the observed NIRCam fluxes (black points with errors), as indicated by inspection and the high \chisq\ (left panel).\footnote{In this case there is no contributed star formation at $z=11$ because its SED would be moderately blue, so there is no amount that would improve the fit to the moderately age population of the observed SED.} The next epoch up shows the best fit at $z=10$ and demonstrates a key feature of \SEDz*: the previous $z=12$ and $z=11$ contributions are not simply ``carried over" to $z=10$. Instead, these are recomputed with new mass estimates for the evolved fluxes from $z=12~and~11$ star formation and adding more at $z=10$.  In other words, the amplitude of star formation at each previous epoch, and the ``present" epoch, are vectors that the NNLS weighs and combines to find the lowest \chisq. The fit at $z=10$ to the observed SED is a bit better, and this improvement continues up the sequence of epochs as star formation is added at $z=9$, $8$, and $7$. The best fit -- the lowest \chisq\ -- that sets the redshift of the SED, derives from contributions from most epochs. This contrasts with the situation for bursty SFHs, discussed in the next section. 

The other two cases shown here, GLASS-3665 and GLASS-3459 show the same behavior: the good fit of the SED at the mimimum \chisq\ develops from $z=11$ as addition epochs of star formation are added. These three examples are typical of the full sample of eight galaxies shown in Fig. 3, and this ``continuing growth" behavior played no role with their inclusion in our sample, and indeed was not recognized before this `evolution-of-the-SED' experiment.  This is our primary evidence that the long, continuous SFHs in Fig. 3 are genuine: achieving the best fit \emph{requires} the addition of stellar populations over a wide redshift range.  

\begin{figure*}
\includegraphics[ scale = 0.75]{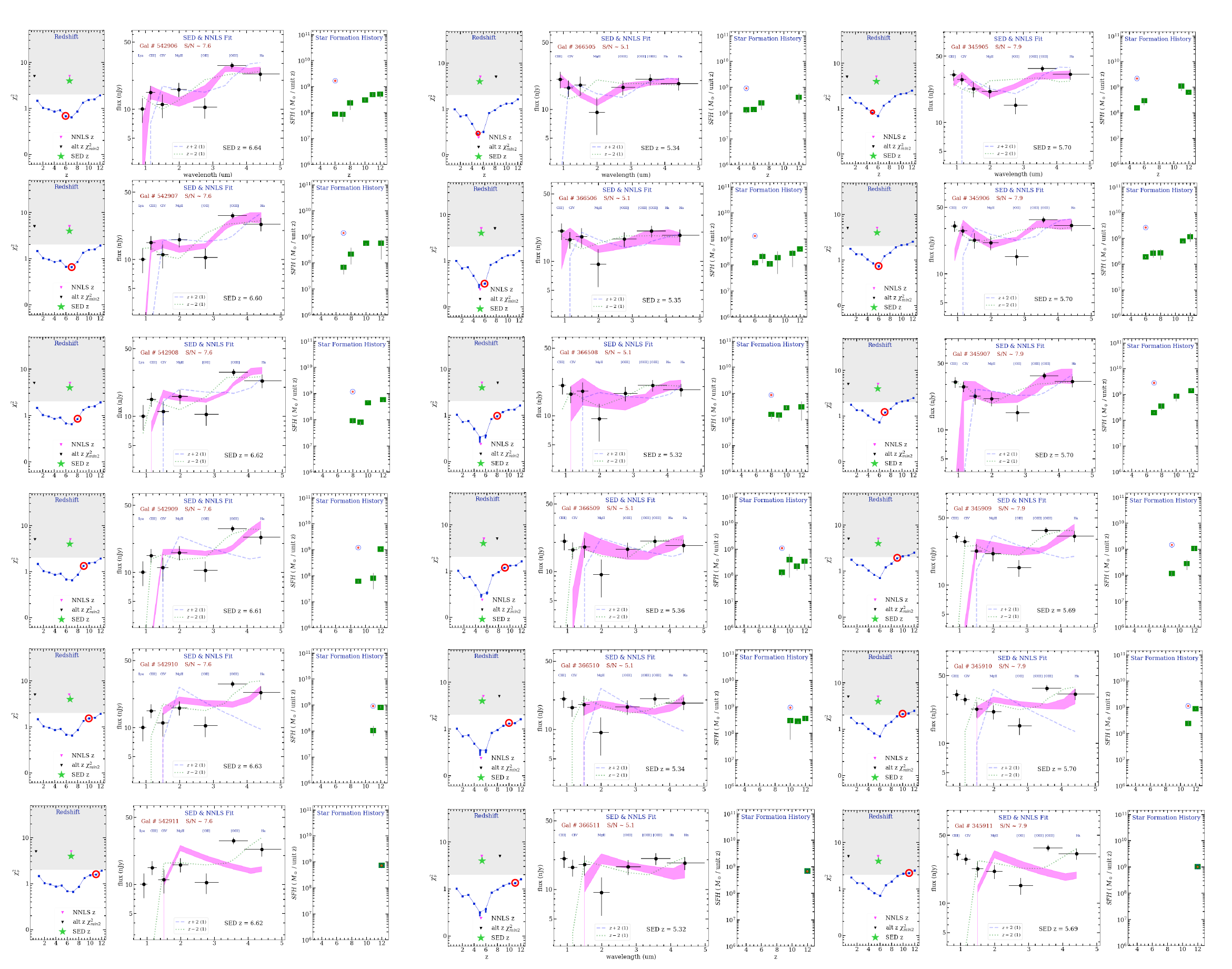}
\caption{The ``growth history" of 3 of the 8 galaxies in Fig. 3, showing a long, mostly continuous star formation from $z=12$.  These three -- GLASS-5429, GLASS-3665,and GLASS-3459 -- are representative of all eight. We plot the operation of \SEDz* epoch-by-epoch to show how contributions from each stellar population are calculated by NNLS, combining stellar population templates and steadily improving \chisq\ -- from the poor fit of the observed SED using star formation only from the earliest epochs (starting at the bottom with $z=11$) to the excellent fit of the SED at the top of each of the three columns (OE).  Panels are labeled by `galaxy number + epoch, for example, `G366508' is the best-fit SED produced by adding stellar templates up from $z=12$ to $z=8$. The red ring in each \chisq\ plot denotes the epoch of that frame. For GLASS-5429 the minimum \chisq\ = 0.8 is reached at an OE of $z=5$. 
}
\end{figure*}
\section{Can \emph{SEDz*} mistake bursts for long SFHs?}
In the previous section we used SFH simulations to show that the long SFHs we have found are, in fact, reproduced by star formation over most of the epochs $z=12$ to $z=5$. Of course, some of these might be better described as a series of 4-5 bursts that with significant drops in star formation in between. Since \SEDz*\ cannot resolve such star formation episodes, a better way to answer this question could be to use a large sample to study the distribution of bursty histories to see if there is a continuum of SFHs that varies between 1 or 2 bursts from $z=12-5$ and the long histories just discussed.  Lacking such a sample at this time, it is useful to ask if well-separated bursts can mimic long histories, at least to the extent that the most or all of the examples in Fig. 3 could be `noisy' versions of bursts.  We investigate this question in this section.

We rigged the \SEDz*\ simulation program to produce random combinations of bursts with mass between $10^{8.5}$ and  $10^{9.5}$\Msun, redshifts z = 8, 9, 10, 11, or 12, and observed at z = 5, 6, or 7. Fifty such combinations were made; the first 28, which are representative, are shown in Fig. 8. For each combination there are a pair of \SEDz*\ plots:  the burst parameters at the chosen epoch (top) and a companion SFH at the chosen OE (bottom).\footnote{The attentive reader will notice that these \chisq\ fits are extraordinarily sharp -- this is an artifact of using the same templates to compose a SFH and for regenerating the SFH, in contrast to the situation with real data where stellar populations are compared to templates.}  Scanning over these examples it is readily seen that, at the redshift of the burst, the SFH is usually a single epoch or two adjacent epochs of star formation, and the observed starburst from a much later time is unresolved and typically splits into several epochs straddling the input epoch of the burst.  At OE the SED is always steep for these: there is no star formation that extends to OE at a significant level, so the relatively old population dominates.

\begin{figure*}
\includegraphics[ scale = 1.2]{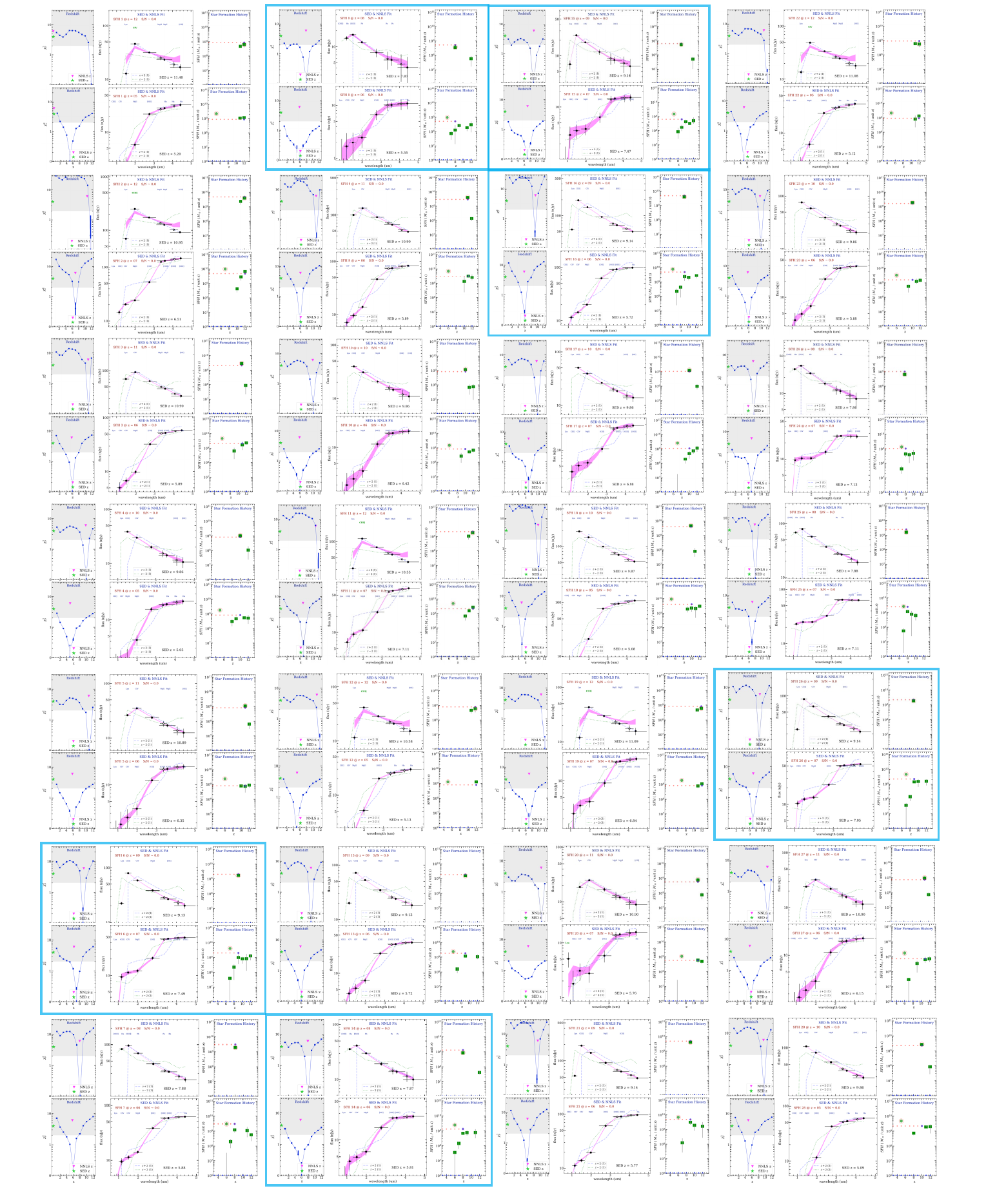}
\caption{28 out of 50 simulations of bursts at $10^{8.5}$ and  $10^{9.5}$\Msun, redshifts of z = 8, 9, 10, 11, or 12, and observed at z = 5, 6, or 7.  For each pair, the \SEDz*\ plot of the bursts epoch (top) is compared to the OE (bottom). The SFH of the burst is usually resolved as 1 or 2 epochs at the burst redshift, and commonly this spreads to an unresolved 3 or 4 epochs around the burst redshift when viewed later, at OE.   However, 6 of the 2; 8 examples (enclosed in a blue box) do show histories that appear to continue down almost to OE -- presenting the appearance of a long SFH.  But, in contrast to the \emph{observed} long SFHs of Fig. 3, a burst SED seen at OE is always red: it is an old stellar population lacking later star formation to moderate its color -- see Fig 9. (Note: In `simulation' plots, total mass is marked by a green-dot/red-ring.)}

\end{figure*}

However, there are 6 cases out of 28 where we see something like the long SFHs of Fig. 3, simulated in the previous section (examples enclosed in a blue box).  We extract these for comparison with 6 of the observed SFHs in Fig. 9.  We see that, although superficially like the long SFHs on the right, there are differences.  Simulations 6, 14, 16, and 26 all have flat SFHs around the burst epoch (the `resolution' problem), with two rapidly falling values -- with large error bars -- to lower redshift. GLASS 2199 on the right is the closest to these in steepness, but its longer history is distinguishable. Simulations 8 and 15 are most like the right-hand examples, which would make this possibility a $<10$\% occurrence.  In point of fact, there is another difference that separates even these mistaken burst SFHs from the examples in Fig. 3: their SEDs are all considerably redder (with the exception of GLASS 2199) -- as expected, the SED of an old burst is not likely to mimic a SFH that has a significant proportion of relatively younger stars.

We conclude from this test that  a small fraction of strong bursts at higher redshift can appear to stretch to longer histories when viewed at $z= 5-7$, but they could only account for at most 1 or 2 of the long SFHs we have identified, and because of the difference in the SED color, even these should be distinguishable.

Reviewing these tests: Figure 6 shows that we can simulate SFHs of the kinds seen in this paper and recover them with sufficient fidelity to distinguish one type from another.  Figure 7 shows that the long SFHs we observe extending to lower z are in fact resolved by \SEDz*\ into many epochs of star formation from $z=12$ to OE. Figures 8 and 9 show through simulations of $8<z<12$ bursts are unlikely to be mistaken for long SFHs.  Even if a burst cannot be pinned down to a certain redshift and duration when viewed from $5<z<7$, their derived SFHs do not contaminate the sample of the other histories we have shown -- they are identifiable as likely bursts in most cases.  The ability of \SEDz*\ to leverage stellar populations whose light is dominated by A-stars to investigate the variety of SFHs at $z>5$ is confirmed.

\begin{figure*}
\includegraphics[ scale = 0.9]{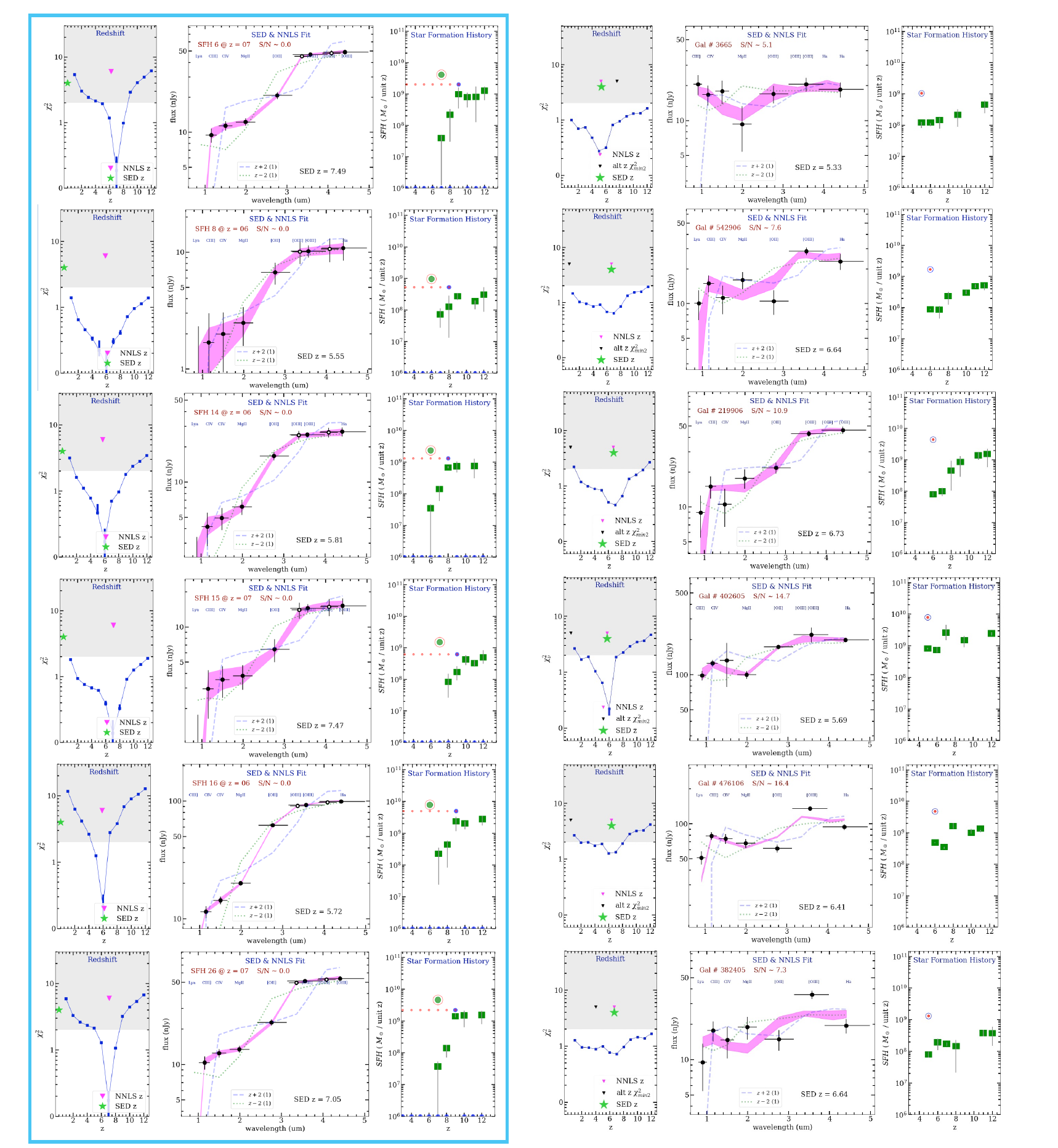}
\caption{Left: the six best examples of \emph{simulated} higher-redshift bursts that might be mistaken for longer SFHs. (Total mass is marked by a green dot/red ring.) Right: Six \emph{observed} examples from Fig. 3, for comparison. (Total mass is a red dot with the blue ring.) In addition to being a small fraction of the burst simulations, the impersonators are distinguishable by their flat-starts/steep-falloffs in star formation in contrast to the steady declines seen in the `observed' sample, and by their steeper SEDs, that come naturally from the domination of older stellar populations.}
\end{figure*}



\bibliography{GLASS}{}
\bibliographystyle{aasjournal}





\end{document}